# A hypergraph model shows the carbon reduction potential of effective space use in housing

Ramon Elias Weber*[a], Caitlin Mueller[a], Christoph Reinhart[a]

[a] Building Technology Program, Department of Architecture, Massachusetts Institute of Technology, 77 Massachusetts Avenue, Cambridge, MA 02139, USA, *reweber@mit.edu

## 1. Abstract

Humans spend over 90% of their time in buildings which account for 40% of anthropogenic greenhouse gas (GHG) emissions, making buildings the leading cause of climate change. To incentivize more sustainable construction, building codes are used to enforce indoor comfort standards and maximum energy use. However, they currently only reward energy efficiency measures such as equipment or envelope upgrades and disregard the actual spatial configuration and usage. Using a new hypergraph model that encodes building floorplan organization and facilitates automatic geometry creation, we demonstrate that space efficiency outperforms envelope upgrades in terms of operational carbon emissions in 72%, 61% and 33% of surveyed buildings in Zurich, New York, and Singapore. Automatically generated floorplans for a case study in Zurich further increase access to daylight by up to 24%, revealing that auto-generated floorplans have the potential to improve the quality of residential spaces in terms of environmental performance and access to daylight.

## 2. Introduction

Current estimates predict that the global built area may grow by a fantastic 250 billion square meters until 2050 to house a growing and more prosperous population [1], [2]. Such estimates are necessarily extrapolations from current building practices. While many decades of building science research and practice have enabled design teams across the world to precisely predict carbon reduction savings that can be attained through any number of upgrades for building operation [3], [4], [5], [6] and materials [7], [8], surprisingly little attention has been paid to space evaluation methods. The ubiquitous energy use intensity metric, defined as energy use per conditioned floor area, has become the de facto benchmark for high performance buildings, leading to sometimes absurd situations where over-sized single-family homes with rooftop photovoltaics are hailed as beacons of sustainability despite of their significant material and space use per occupant.

Given that energy use roughly scales with building size, reducing the floor area per apartment unit while maintaining good indoor environmental conditions offers a complementary path towards a net zero building stock. Traditional architectural design workflows are unsuitable for this type of exploration since they rely on a human to manually draw interior walls while considering a plethora of architectural, safety, and ADA requirements [9]. In large projects, this step tends to be delegated to junior designers who are given references from previous projects to follow and then draw interior walls in reference to the building's structural grid. The position of these interior partitions obviously impacts access to daylight, thermal comfort, and view to the outside. In many cases, overall performance could probably be improved through more extensive design explorations. However, that barely happens due to the additional time effort and technical sophistication required to conduct this type of analysis.

While the construction industry has long shied away from quantitatively evaluating space use, the urgency of the climate crisis along with a shortage of architects to meet the global housing demand has led to some, mostly developer-driven and proprietary attempts to automatically generate floor plans especially for apartment buildings [10]. Most current implementations are linked to financial costing models, evaluating multiple ways to divide a building footprint into a desired number of apartment units [11], [12], [13]. Current approaches for within-unit room divisions are an active area of computer graphics research but are not currently used in the architecture industry due to a wide range of limitations: Being only able to represent rectangular [14] or orthogonal boundary conditions [15], or responding to only either topological or spatial or boundary constraints [16], [17], [18], [19], [20]. On a technical level, ML-based models create neural networks that relate the geometric graph structures from room walls to an adjacency graph (vector [21], [22] or pixel based [23]) or use reinforcement learning to subdivide a space [24]. This results in a linear, one-sided generation process, where a room adjacency graph is converted into a visually real and geometrically valid floor plan. Inherently these processes do not allow for exact specifications of room sizes, boundary conditions or further geometric manipulation of parts of the final output, as needed in architectural design. Furthermore, implicit geometric relationships are difficult to train, and floor plan training data is sparce, scarce and unvetted; thus, such approaches can neither guarantee architectural quality nor environmental performance [10].

Instead, we present the hypergraph, a generalizable shape generator and descriptor for floor plans. The hypergraph represents key characteristics of the shape divisions of any given floor plan layout, enabling both the mapping and benchmarking of suitable, high performing floor plans, as well as their automatic generation. A hypergraph is created from existing building floor plans and can be applied to new conditions. This allows for translating cultural conventions and practices into new designs, and a fully transparent source attribution. We introduce a spatial analysis workflow to minimize "excess space" while retaining the same spatial functionality of a given floor plan. The concept of excess space is based on the notion that a room with a certain program, say a bedroom, has minimum functional requirements in terms of furniture (bed, dresser, cabinet) and space around that furniture that supports its proper use. Areas beyond those functional requirements are then defined as excess. Furthermore, an automatic integration of environmental analysis methods of energy and daylight allows us to benchmark high performance designs and helps maximizing occupant comfort conditions.

## 3. Main

**The hypergraph framework is a graph-based representation of an architectural floor plan**. In the architectural design process of residential buildings, the fitting process of apartment units, as well as the internal subdivision inside the units to create a floor plan remains a process that is typically performed manually by an architect. The hypergraph aims to computationally execute this two-dimensional design process by providing a unique mapping that either divides a building outline into apartments or an apartment into rooms. This mapping can be applied to any building outline and is stored as a graph-based representation. In this paper we will apply the hypergraphs to subdivide residential apartment units into rooms. To generate a hypergraph, key components of a typical architectural representation of a floor plan (Figure 1a) are analyzed to extract different rooms with their specified program and the connectivity between rooms. A binary space partition tree [25], is constructed to represent the geometric subdivision into different rooms, where the outer most nodes are the actual rooms of the apartment, color coded by program. An undirected access graph represents the connectivity between rooms and through that defines the spatial organization of the floor plan (Figure 1b). The resulting hypergraph is a combination of the two subgraphs, the binary space partition tree and the access graph, with nodes representing rooms and edges representing spatial subdivision or access (Figure 1c) *(see Methods: 4.3 Creation of Hypergraphs).* The binary space partition tree can be generated from any non-convex and most convex spatial configurations which allowed us to represent the real-world floor plans we encountered (Supplementary Fig 8-10). Given the same boundary condition, the hypergraph constitutes a bijective mapping that results in the same floor plan and vice versa. The same hypergraph can be applied to a variety of boundary polygons that will result in a unique floor plan for each boundary condition (Figure 1d) and the other way around (Figure 1e).

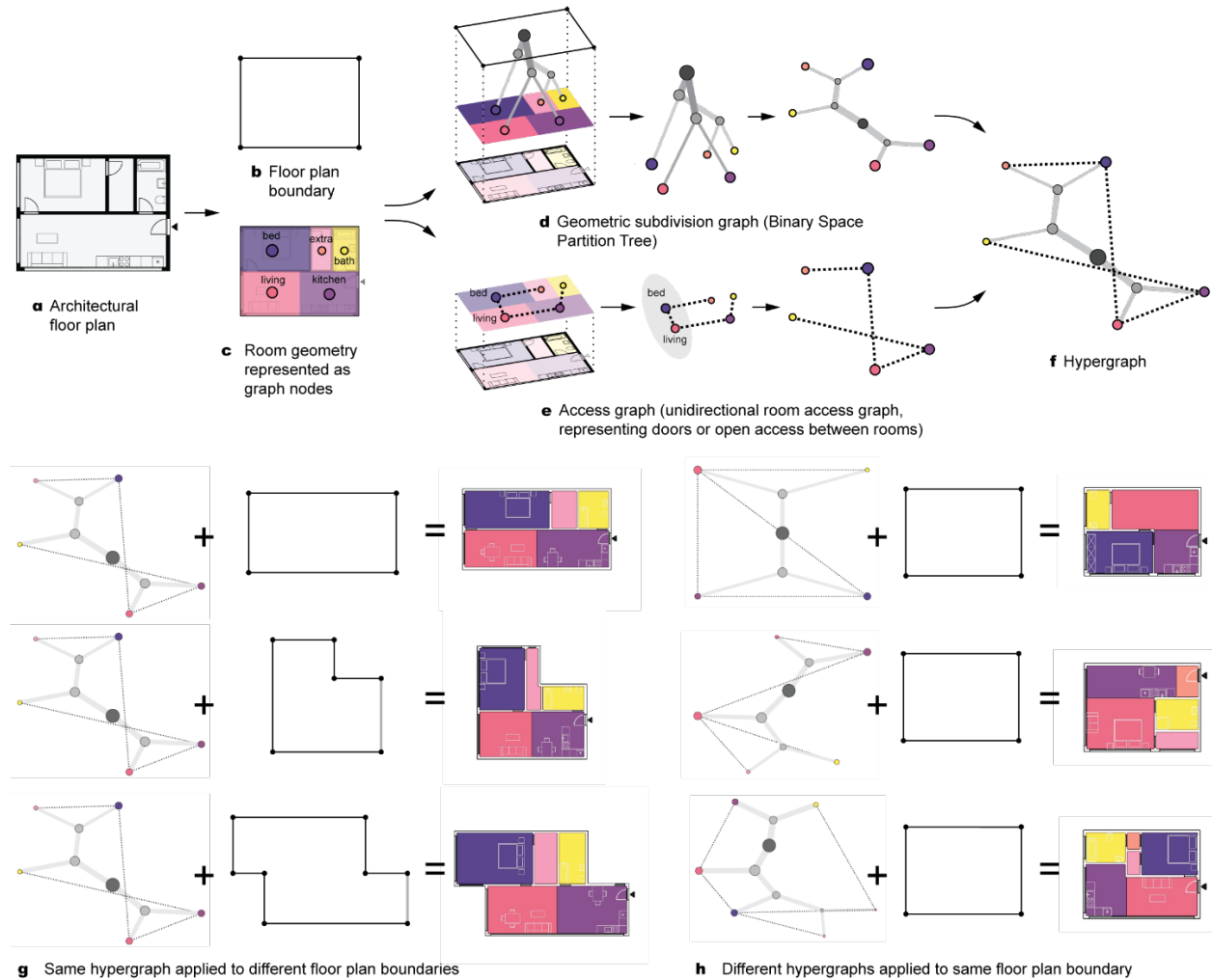

*Figure 1: The hypergraph is generated from an architectural floorplan (a) that is converted into a boundary (b) and programmatic zones (c) that are translated into the graph nodes. Geometric subdivision of boundary into the rooms is computed with a binary space partition tree (d). The access between rooms is represented as a unidirectional room access graph (e). Combined they result in a hypergraph (f). A hypergraph can be applied to different floor plan boundary (g) to create floor plans with a similar typology. Different hypergraphs can be applied to the same floor plan boundary to create floor plans with different internal configuration (h).*

**The hypergraph framework can be used to create and test different floor plans and assess them from a spatial and environmental perspective.** To describe a whole building, we can apply the hypergraphs to an apartment boundary, generating detailed floor plans for each apartment unit. A fitting procedure is shown in detail in Figure 2, where an apartment boundary polygon (Figure 2a) is subdivided by a library of different hypergraphs (Figure 2b) to create different internal apartment configurations (Figure 2c). We then use the apartment boundary polygon and its orientation towards the building circulation to filter floor plans with similar orientations and façade to adiabatic wall ratios. The hypergraph method removes the need for drawing floor plans and preparing geometry for different simulation procedures and allows the very complex structure of a floor plan to be described by

quantifiable and searchable key parameters and graphs. To filter geometrically valid but spatially inadequate outputs a series of heuristics filter and rank feasible results *(see Methods: 4.7 Apartment Validity Heuristic)*. With this, we can generate architecturally feasible floor plans where rooms have an aspect ratio and size that makes them usable for their specified use, have access to a façade and are configured in a way that allows access within the apartment as well as to the building's circulation. For assessing the spatial validity of a floor plan, we propose an automatic version of the spatial scoring system developed by the City of Berlin's public housing provider [26]. Using automatic placement of furniture blocks we can assess if rooms are large enough to result in livable spaces and compare the overall area to reference floor plans with the same occupancy (Figure 2d) *(see Methods: 4.9 Furniture Placement)*.

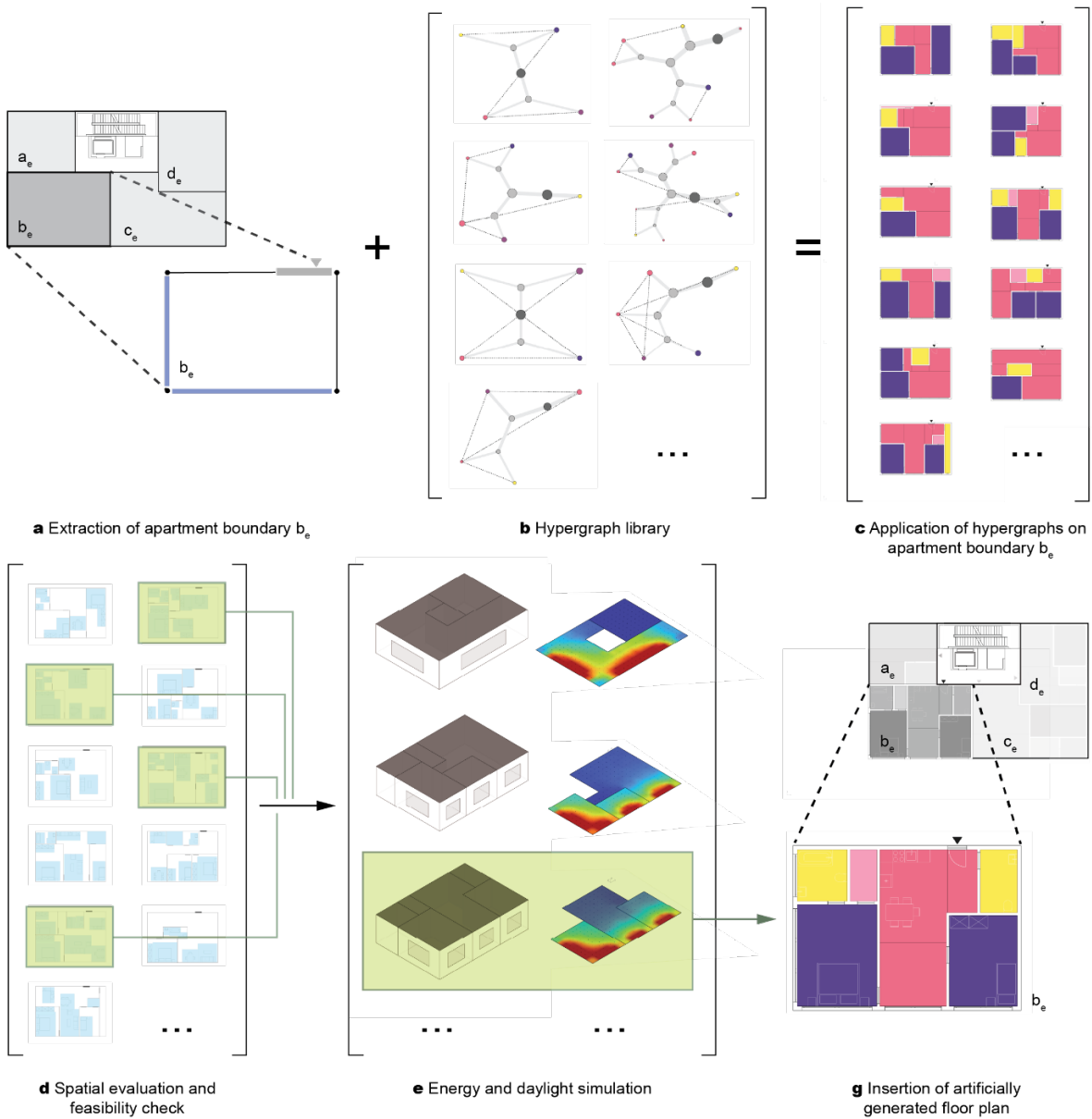

**a** Extraction of apartment boundary $b_e$  
**b** Hypergraph library  
**c** Application of hypergraphs on apartment boundary $b_e$  
**d** Spatial evaluation and feasibility check  
**e** Energy and daylight simulation  
**g** Insertion of artificially generated floor plan  

*Figure 2: Steps for fitting an apartment using the hypergraph method. An apartment boundary is extracted from a building (a) and combined with a library of hypergraphs (b). The applied hypergraphs generate different internal subdivisions for the apartment boundary (c). A spatial evaluation using placement of furniture, accessibility and room geometry is performed to filter*

*feasible solutions (d). An energy (e) and daylight analysis (f) is performed to evaluate the resulting floor plan and a chosen plan is inserted into the building (g).*

To estimate daylight and energy performance, the selected floor plans are automatically converted into a of a 3D model with walls and windows that creates a building energy model of the apartment. Building energy models are heat-transfer and mass flow simulations that are industry standard for energy use predictions [27]. Furthermore, we calculate daylight access in the apartment through assessing the Spatial Daylight Autonomy (sDA), a metric for interior spaces that, through a yearly illuminance simulation with physics-based raytracing and local weather data [28], predicts the percentage of hours per year, when a minimum light level of 300 lux can be achieved with daylight (Figure 2e). While whole building energy models typically do not have the geometric resolution of single rooms, the models generated with the hypergraph method will allow more detailed energy performance analysis that can capture effects of airflow and natural ventilation for more accurate predictions. Furthermore, daylight predictions do require high geometric detail, as the internal configuration of a floor plan will determine how light gets obstructed inside an apartment.

**Hypergraphs allow the spatial characterization and comparison of floor plans.** To show the spatial analysis potential of the hypergraph framework we created a dataset of residential floor plans from around the globe (*see Methods 4.1. Residential building floor plan repository*). In order to characterize differences between cities we compare a representative subset of floor plans from three different cities: Zurich, New York and Singapore. Contrary to explicit representations with Euclidian geometry or pixel-based representations, the hypergraph encodes relative spatial relationships in addition to geometric properties. This allows the mapping of spatial and typological similarities between floor plans that have different boundary conditions. Based on the number of rooms, subdivision graphs have a variety of sizes, requiring comparison functions to work with matrices of different dimensions. For comparison of different hypergraphs, we compare the subdivision matrix (derived from the spatial subdivision graph) separately from the access matrix (derived from the room access graph).

We can describe the overall configuration and complexity of a hypergraph through the number and degree of access and subdivision edges, normalized by the number of rooms. Using a principal component analysis (PCA) of key attributes (see Supplemental Figure 9) we can demonstrate that the hypergraph method can be used to distinguish and group similar floor plans according to size and occupancy, as well as compactness (Figure 3a). Hypergraphs with lower graph complexity, corresponding to smaller size and occupancy are grouped in the left around the x-axis, while more complex configurations are grouped towards the right. This creates opportunities to quantify spatial differences of apartments across cities, such as simpler spatial properties of apartments in New York, when compared to small-scale more complex floor plans with higher hypergraph subdivisions in Singapore. The hypergraph allows to show and encapsulate architectural differences and investigate of spatial configurations that are encoded in local cultural architectural practices, prevalent construction techniques, building code and climate.

**The geometric resolution of the hypergraph generated models allows for new insights into building performance**. An automated spatial and environmental analysis allows us to capture differences in daylighting across the three cities. The results of our simulations support qualitative architectural observations, that residential apartments in Zurich and Singapore have more access to daylight and are mostly daylit from different sides, while apartments in New York in larger buildings have less daylight access (Figure 3b). To study the energy efficiency of different apartment geometries, we derive two building energy models for each apartment with a standard and a high-performance building envelope *(see Methods 4.8. Environmental evaluation workflow)*. The difference in energy use of the two models shows the energy savings from upgrading the building envelope. We conduct an automated spatial analysis to assess if a floor plan is usable and how its area compares to the minimum size requirements for its occupancy *(see Methods 4.7-4.10)*. With this we can quantify the unused space of a floor plan and with it the excess emissions associated with heating or cooling. Floor plans that are too large in area or have large 'unusable' circulation areas are penalized. A comparison shows how excess emissions from unused space can be significantly higher than savings from building energy upgrades, assuming that the size of the apartment could be reduced until no excess space remains. We find unused space to be more impactful than envelope upgrades, especially in the more temperate climate in Zurich (71.6 %), while the opposite is found in hot and humid Singapore (33.0 %) where floor plans are already very compact and envelope performance is crucial due to the climate. In New York a balance from both measures yields best results (61.0 %) (Figure 3c). This means that in the case of new construction, apartments that are closer to the minimal spatial requirements with less excess space will have significantly lower energy use, even when constructed with lower envelope standards.

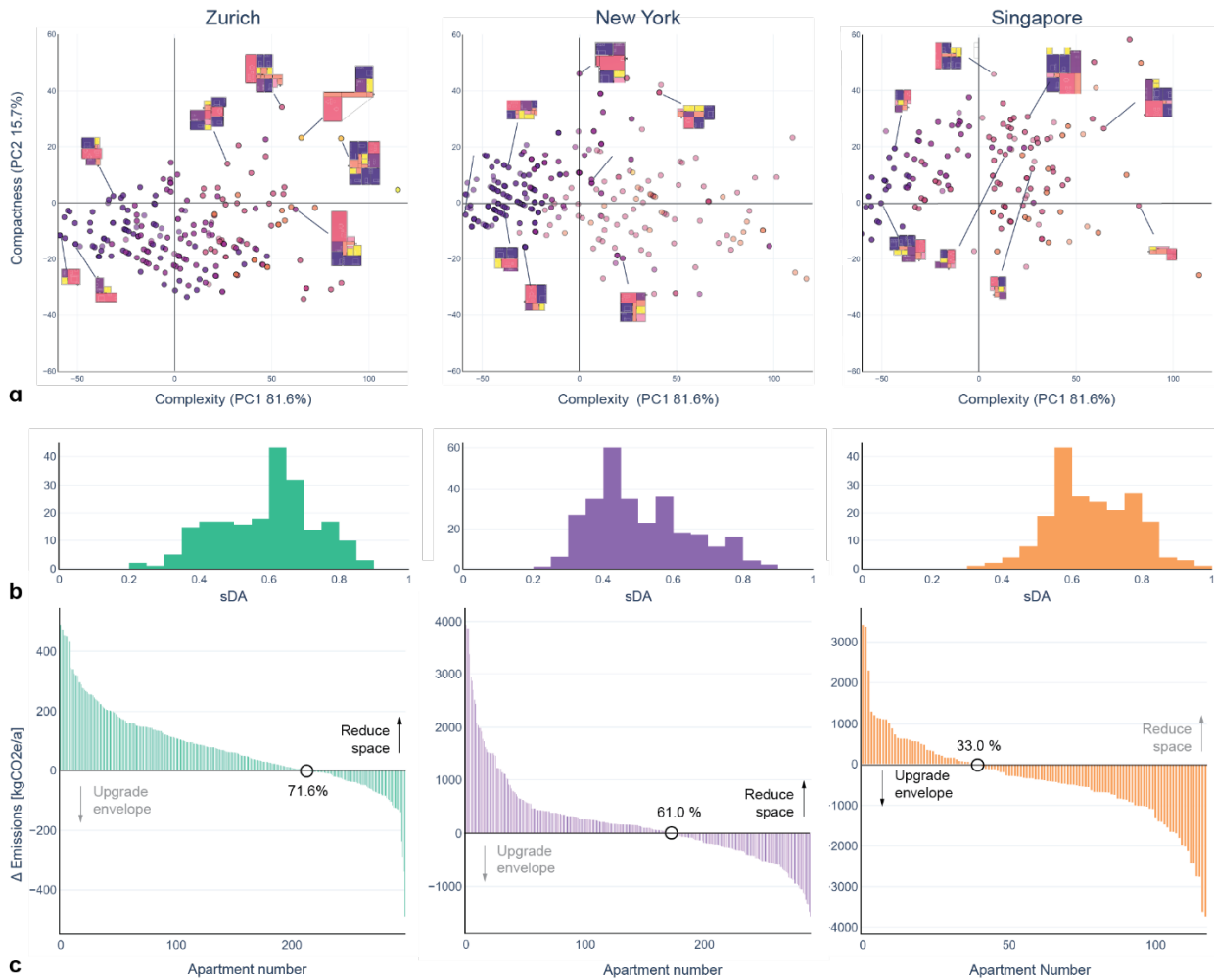

*Figure 3:* Mapping of all floor plans of Zurich, New York and Singapore with their graph structure (a) and sDA performance (b). Comparison of emission from excess space compared with envelope upgrades (c) from the calibrated dataset *(see 5.1. Residential building floor plan repository)*.

**Artificially generated floor plans using the hypergraph framework can match and outperform real-world buildings.** Apart from analyzing existing floorplans, hypergraphs can also be used to generate new floorplans. We use the hypergraphs of all collected floor plans and test fit them

automatically into real-world boundary geometries of residential buildings from Zurich, New York, and Singapore. A sample building for each city that was fitted is shown in Figure 4a (for all reference buildings see Supplemental Figures 10-12). Depending on the apartment boundary there are different numbers of valid apartment subdivisions possible, and the hypergraph fitting method was able to propose alternative apartment layouts inside the real-world buildings. The apartments created through the hypergraph fitting method were then assessed for daylight to compare their sDA. Even though not all of the artificially generated floorplans would be spatially desirable, the aggregated results of the simulation could be used to predict the daylight performance of a building (Figure 4b). When comparing the sDA performance of the artificially created apartments, the third quartile of results is within a 20% range or better than the real-world floor plans for sDA and in 5% of cases (Zurich) outperformed the reference floor plan by up to 24% (5.8% in New York by up to 16%, 0.4% Singapore by up to 10%). Furthermore, a more detailed qualitative analysis of example floor plans that performed in the upper percentiles of the performance ranges reveal significant opportunities for design of new buildings: different spatial configurations that increase the daylight significantly, alternative spatial configurations that retain daylight performance, opportunities for adding additional rooms, which indicates the chosen number of rooms might be too small, or reducing the number of rooms, which indicates that a floor plan might be too tightly fitted (Figure 4c).

## *4. Discussion*

In summary, we have demonstrated how the hypergraph framework, as a bijective mapping procedure for creating and representing apartment floor plans can be used to describe spaces across the world. To our knowledge, the hypergraph is the first method that can generally describe apartment geometries and can translate architectural geometry into graph-based representations and vice versa. We could show how the method can be used for mapping and comparing different apartments and propose alternative solutions for existing buildings. These assessments will impact retrofit decisions and regulations on a policy and city planning level, allowing us to better understand and shape, dense urban environments. Secondly, the automated testing and generation of multiple design options will create opportunities for better design of new buildings, from providing ideas for designers to having quality controls that will predict achievable daylight levels and energy performance for a given context. It could allow for new types of software that would allow self-building and design for communities that cannot afford professional architects, while ensuring that the automatically generated buildings have architectural precedents that promote healthy and sustainable spaces.

Currently, emissions from buildings vary greatly [29] and our method shows new pathways for helping architects converge the energy performance and spatial requirements on an urban level with the comfort and needs of building inhabitants. The automated nature of the procedure lowers the barrier for environmental simulations of all buildings, which is key in enabling sustainable building design across the globe. We can show how in the design of a building the spatial configuration is more important than specifications of building envelopes when it comes to building energy usage. Both in the surveyed reference floor plans, as well as in our artificially generated ones it would have been more favorable in terms of total carbon emissions to build less space, instead of more high performance. With this we can prove that space sufficiency could be used as a highly impactful factor carbon mitigation strategy of cities with large impacts on future building energy policy. These insights should guide the standards and building codes of cities in the future. An overhaul of current environmental certifications is needed [30], such as the cost balance method in ASHRAE 90.1 [31] or LEED [32] standards that currently do not award spatial efficiency and compactness. Given the prevalence of energy use intensity, smaller spaces are penalized because of a higher "equipment" per floor area ratios. Contrary to current energy codes that specify performance requirements our results show great potential in savings through spatial efficiency measures, the thoughtful planning and design of buildings and with it the possibility to include spatial metric for designing of building energy sufficiency [33].

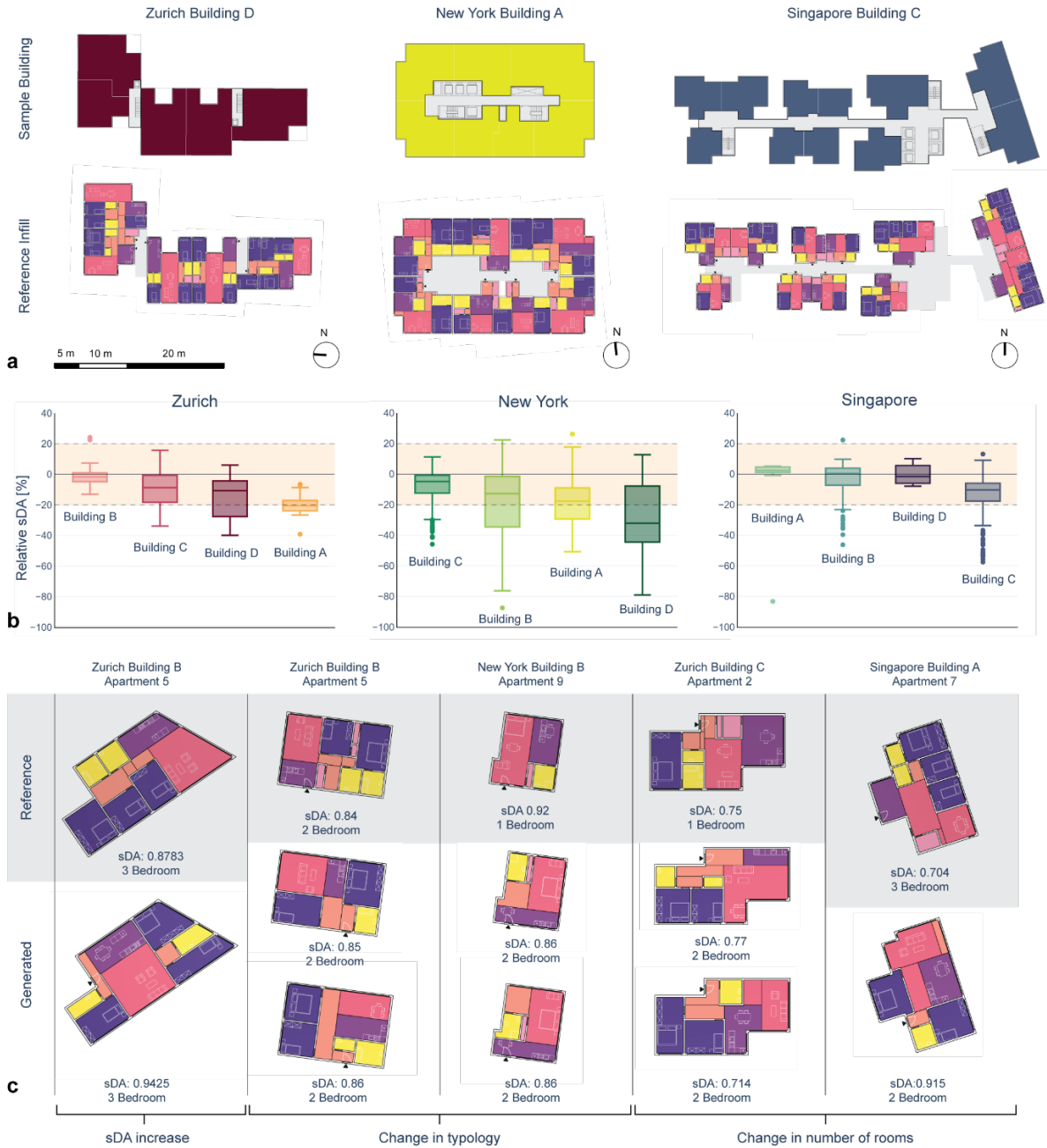

*Figure 4: Three sample buildings with reference floor plans (a) that are being replaced by hypergraph generated floorplans (all buildings are defined in Supplemental Figures 10-12). The relative sDA performance of all successful floor plans with equal or more rooms (b) and single artificially generated floor plan examples (c) highlighting different opportunities.*

While the paper shows substantial promise for using automated floor plans for lowering building energy use, the authors acknowledge that there are important questions of ownership when using an automated procedure that is based on precedent designs. If clearly vetted floorplans are in the public domain, or are generated in-house by an architecture firm, reproducing geometric configurations will be highly beneficial to increase the speed of design workflows. Questions of intellectual ownership will arise that will have to be addressed by legislators, which ties into the existing debate around large language models [34] and generative AI [35]. However, a key difference with the hypergraph approach is a clear source attribution of each graph and the possibility to explicitly map differences and similarities to existing designs.

The fully automated method can be used to create architecturally valid apartments that can be combined with environmental performance analysis to evaluate and automatically generate culturally relevant, high performing buildings. By utilizing architecturally vetted reference designs and heuristic procedures that respond to local requirements, the hypergraph method can produce high quality spaces from virtually any boundary condition. While the method yields geometrically valid options, these designs may not always be of sufficient architectural quality as they depend on the quality of the underlying floorplans in the reference database. However, using only a minimal dataset, we managed to generate artificial solutions that are on par and up to 24% better in daylight performance than the real-world built references. This reveals that our method has great potential to lastingly improve the performance of new construction worldwide. We further see opportunities to apply the method to automated benchmarking of building retrofits including the conversions [36] of some of the currently 20% empty office buildings [37] in the US to residential units [38].

## 5. Methods

### 5.1. Residential building floor plan repository

We assembled a reference library of ~1,444 real world floor plans, combining award winning residential floor plans North American and European context from literature [39], [40] [41] and online databases [42] with residential developer plans. The library contains unique real-world floor plans (and their mirrored geometry). Using publicly available data from real-estate brokers and public housing providers, the dataset represents a small subset of a city's actual apartments. However, we curated the library to encompass a variety of different design, and to represent the prevalent apartment layouts of the cities, with studio apartments to large multi-unit apartment units and across different price ranges from public housing to high-end apartments. From the reference library dataset floorplans are sampled to map the distribution of number of bedrooms of the real-world data surveyed in Singapore [43], New York [44] and Zurich [45] . Please note that because almost 80% of residents in Singapore live in government provided housing [46] that is built in a standardized fashion there is less variety in the building stock. This is reflected in a smaller dataset as in Zurich or New York. We compare the real-world data with our dataset in Figure 5. The full library can be accessed via the supplemental data of this paper.

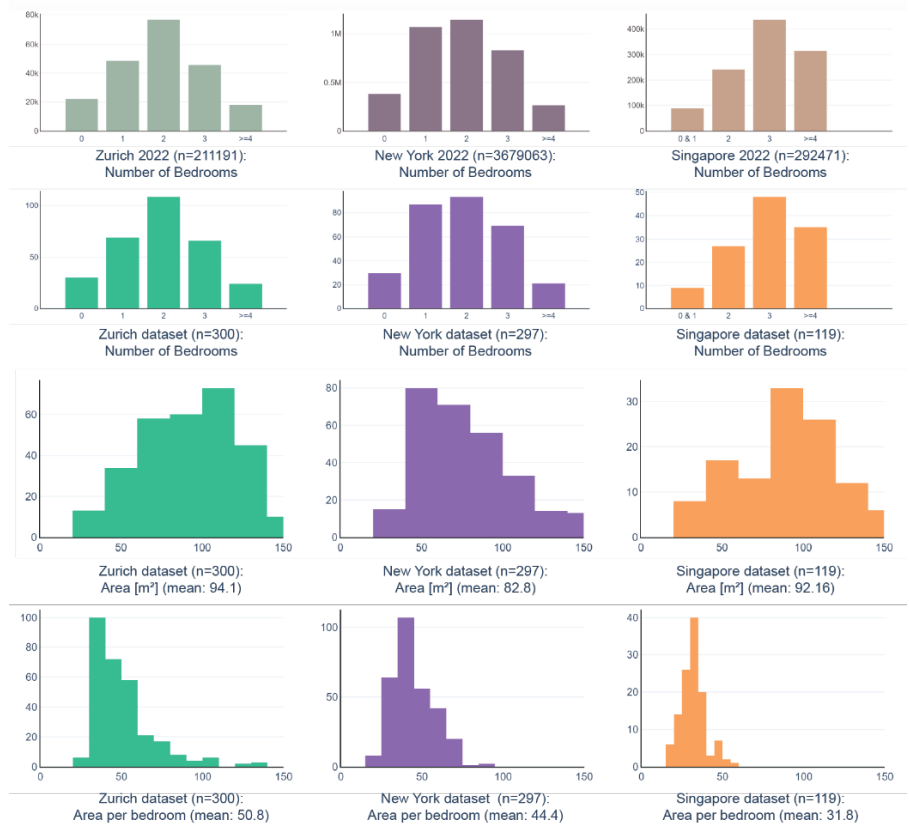

*Figure 5: Distribution of the number of rooms of the cities of Zurich, New York and Singapore, compared with our reference dataset of analyzed floorplans.*

### 5.2. Reference apartment buildings for artificial floor plan insertion

To test the artificial generation of floor plans and application of reference floor plans into new boundary conditions we gathered reference buildings from three different cities. Four buildings from Singapore, Zurich and New York were chosen to qualitatively reflect contemporary residential housing architecture of their city. For reasons of data protection of the residents, as well as the architects, we have anonymized the buildings and refer to them as Building A, B, C, D from their respective city. The building boundaries, as well as the floor plans for Zurich New York and for Singapore, we used to benchmark the artificially generated floor plans can be accessed via the supplemental material (Supp. Figure 8-10).

### 5.3. Creation of hypergraphs

We introduce the hypergraph as a shape descriptor for building floor plans. Graph based data structures have been applied successfully to represent and generate structured data in biology [47], chemistry [48], robotics [49], building structures [50], computer games [51], and urban planning [52]. For the design of building floor plans graph based data structures have been deployed to represent wall lines and adjacency graphs [10]. The presented hypergraphs are a combination between an access graph and a subdivision graph. While previous work used the explicit geometric structure of, for example, a molecule, wall segment or street intersection as a part of a graph, the hypergraph in our case is a combination of explicit geometry through adjacency of specific rooms and implicit geometric representation through the subdivision graph. Both graphs can be accessed and analyzed independently via edge and node type specification in our custom data-format.

The subdivision graph is representing a binary space partition (BSP) [25] tree that simultaneously represents the final geometry, as well as its step-by-step construction. Each node corresponds to an area (or ratio) and a subdivision angle $\alpha$, with (directed) edges connecting the child nodes to the parent node that was subdivided (Figure 6). Our BSP implementation allows for subdivision of polygons with 3 or more boundary vertices and includes convex and (most) nonconvex polygons. In the BSP tree the root node specifies the overall area of the subdivision graph. Subsequent children (of type "subdivision") always have degree 2 and assigned areas, as well as a subdivision angle alpha. Leaf nodes of the subdivision graph have a degree 0 on the subdivision graph and area assigned a programmatic type of either {living, bedroom, kitchen, bath, extra, foyer} and a unique id. The access graph is defined by lists of unique ids in the leaf nodes. Different hypergraphs are illustrated with annotated edges in Figure 6 (and additionally in Supplemental Figure 8). Compared to existing methods, the purely geometric nature of the hypergraph creates a direct relationship between graph and spatial form. It is an explicit and not an iterative or optimization-based process that can be computed in real-time.

The outer most child nodes therefore represent the rooms in the final floor plan, while inner nodes correspond to the intermediate parent areas in the subdivision process. Even though the room adjacencies are defined geometrically through the subdivision the access graph represents the spatial experienced adjacency by (undirected) edges that connect the room nodes (e.g. through a door or an open wall). This dual representation of the internal organization can be captured from any given floor plan boundary. Furthermore, a mapping of both graph nodes of the subdivision and adjacency graphs to the resulting rooms allows for the recording of secondary information, such as room type. The procedure is fully reversible, meaning that a spatial floor plan layout can be encoded in a graph and the same floor plan layout will emerge given a graph and the original boundary polygon. The subdivision method can apply to all nonconvex and most concave polygons with a small convexity score, which allowed the encoding of all real-world floor plans we encountered.

### 5.4. Preprocessing and data preparation of floor plans

The apartment floor plans were sourced as raster images. They were input into the CAD software Rhino where the images were traced and rooms annotated with their respective program, circulation access, façade with lists of lines and room access (doors). We deploy the inverse of the subdivision algorithm to find the corresponding subdivision graph, and the points in the door locations to determine access via the access graph. Both graphs are combined into a hypergraph and stored together with façade, circulation, and boundary lines in a json database.

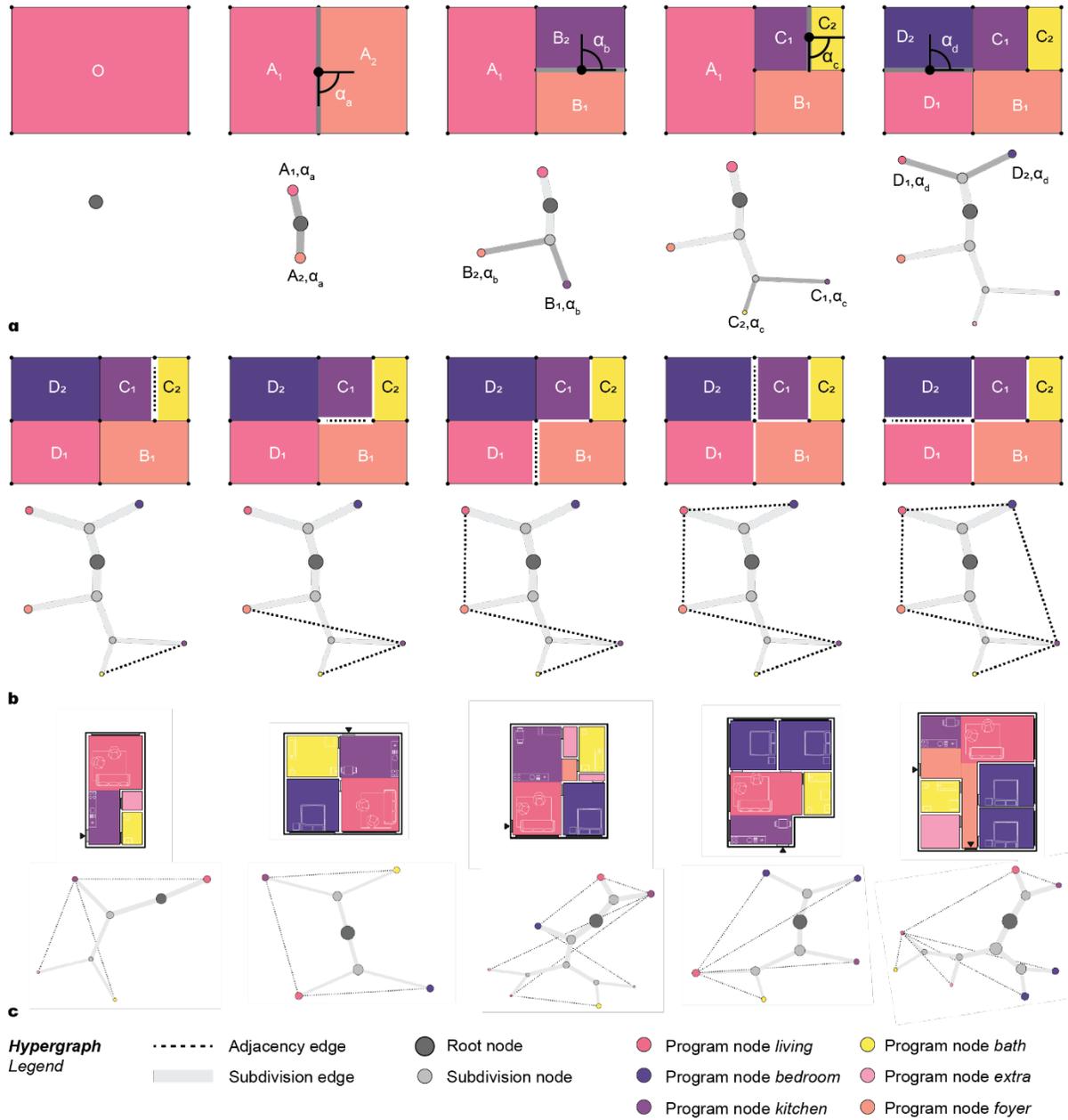

*Figure 6: Step by step generation of the subdivision graph (a) from area O, represented by the grey point (graph root). It is subsequently divided into area A₁ and A₂ with the angle αₐ, A2 is divided into B₁ and B₂ with angle α, B2 is divided into C1 and C2 with angle angle αc and A1 divided into D1 and D2 with angle αd. The access between rooms is converted into a graph (b) where edges connect the room nodes of the subdivision graph that connect (e.g. the rooms that are accessible between one another). Different subdivisions therefore result in different hypergraphs (c).*

## 5.5. Implementation and visualization

We implemented the geometric floor plan placement process in the commercial architectural CAD software Rhino via a custom geometry library in C# through the scripting platform Grasshopper [53]. For this we utilize and extend the functionalities of the open source linear algebra library Math.NET [54] and the 2D polygon clipping and offsetting library Clipper2 [55]. For the visualization and representation of the hypergraphs we convert the graph data structure to a NetworkX [56] graph and visualize it with the force based Kamada-Kawai algorithm [57] that is applied to the nodes of the subdivision graph. The full code of the implementations, as well as sample files showcasing hypergraph creation and analysis, can be accessed via the supplementary software submission.

## 5.6. Limitations of the BSP subdivision graph representation

In the current BSP tree implementation we can represent almost any geometric polygon and subdivision. Even though we were able to represent the studied buildings, there are certain limitations for apartment geometry and configuration that currently cannot be captured in the data format. Failure cases of the subdivision algorithm include highly complex nonconvex boundary geometries, as well as polygonal boundaries with holes. While convex boundary geometries are guaranteed to produce a feasible result, highly convex boundary conditions do not. Typical apartment layouts, and those that we observed in our database fall into this category, however, this is not guaranteed, especially for synthetic datasets. On an architectural level we limited the scope of the current implementation to single story floor plans of multi-unit residential buildings, excluding duplex apartments and single-family homes.

## 5.7. Apartment validity heuristic.

Even though the subdivision algorithm produces a geometrically feasible floor plan, the resulting geometry might not be spatially valid. Different failure cases exist where apartment boundaries are subdivided and produce architecturally infeasible rooms that are inaccessible or don't have access to daylight (Supplemental Figure 2). For creating artificial floor plans that would be further used in a design context, a visual inspection of the results together with placed furniture items that visualize the scale of rooms, proved to be useful. However, for analysis of large-scale datasets automatic procedures are needed to identify feasible results. Since all hypergraphs are created from a geometrically feasible reference floor plan, we can compare the room geometry of the artificially created floor plan with the original reference floor plan. For this comparison to be computationally efficient, we utilize a scoring method that is computed from the perimeter of the room polygons. The perimeter difference score (Equation 1) can be applied to single room polygons (Supplemental Figure 3), as well as whole apartment floor plans (Supp. Figure 4) to determine geometric changes between target and reference. It is a computationally efficient indicator of fit. For more accurate control, more computationally intensive pathway and geometry analysis could be envisioned [48]. Furthermore, we can use the furniture placement algorithm to verify if an apartment is feasible by comparing the minimum required furniture to the placed furniture (Supplemental Figure 7).

$$\delta_p = \left| 1 - \frac{L_{SA} L_B}{L_A L_{SB}} \right|$$

*Equation 1:* Perimeter difference score $\delta_P$, where $L_A$ is the perimeter of polygon A, $L_{SA}$ the perimeter of the square polygon with the same area as A, $L_B$ the perimeter of polygon B, and $L_{SB}$ the perimeter of the square polygon with the same area as B.

## 5.8. Environmental evaluation workflow

The automated workflow was implemented in the commercial architectural CAD software Rhino and its integrated scripting platform Grasshopper [53] where the generated floor plan geometry can be automatically converted to be used by the energy simulation software EnergyPlus [58] and the lighting simulation tool Radiance through the Climate Studio package [59]. The simulations were conducted on a Windows computer with the following specifications: 64 GB Ram, Nvidia GeForce GTX 1080 Graphics card, Intel(R) Core(TM) i7-6700 K @ 4.0 GHz Processor. The full view, daylight and energy simulation required >10s of calculation time per apartment. Settings for the energy simulation of each city and settings for high and standard performing building envelopes are listed in Table 1. For each apartment we calculate the Energy Use Intensity (EUI) in kWh/m$^2$/yr for both a standard and high-performance building envelope. To only compare building geometry related factors, we keep the HVAC system the same, even though in a standard building energy retrofit a more efficient HVAC system could be installed. To calculate the spatial daylight autonomy (sDA, indicating the fraction of space with more than 300 lux of daylight on average) we only look at specific rooms in an apartment that require daylight, excluding bathrooms and extra (storage) space (Supplemental Figure 5). Furthermore, we create an sDA score of each apartment by weighing the area of each room (Equation 2).

*Table 1: Energy simulation settings for climate studio and energy plus*

| City | New York | Singapore | Zurich |
|---|---|---|---|
| *Energy Zone Template* | Ashrae 90.1 – Climate Zone 4 | Ashrae 90.1 – Climate Zone 1 | SIA 2024 |
| *Weather File* | USA_NY_New.York-LaGuardia.AP.725030_TMYx.2004-2018.epw | SGP_SG_Changi.Intl.AP.486980_TMYx.2004-2018.epw | CHE_ZH_Zurich.Fluntern.066600_TMYx.2004-2018.epw |
| *Grid Carbon Intensity (kg/kWh)* | 0.55 [51] | 0.4057 [52] | 0.128 [53] |
| *Energy Template* | Ashrae 90.1 – Climate Zone 4 | Ashrae 90.1 – Climate Zone 1 | SIA 2024 |
| **Standard Envelope** | | | |
| *U value (W/m$^2$K)* | 0.3 | | |
| *Window to Wall Ratio (WWR)* | 0.6 | | |
| *Window* | DoublePaneClr | | |
| *HVAC* | Standard Electric HP COP [3,3] | | |
| **High-Performance Envelope** | | | |
| *U value (W/m$^2$K)* | 0.1 | | |
| *Window to Wall Ratio (WWR)* | 0.6 | | |
| *Window* | Triple Pane LoE | | |
| *HVAC* | Standard Electric HP COP [3,3] | | |

$$d_{tot} = \frac{\sum_{i=1}^{n}(d_i * A_i)}{\sum_{i=1}^{n}(A_i)}$$

*Equation 2:* To get the apartments overall daylight score $d_{tot}$ we multiply the area of each daylit space with its sDA value from our radiance simulation and divide it by the sum of the area of all daylit spaces.

## 5.9. Furniture placement

To spatially evaluate a floor plan, we test fit the layout with furniture. In the computer graphics discipline furniture placement algorithms have been widely explored using machine learning and procedural techniques [16], [60], [61], [62]. The use of furniture blocks to test spatial feasibility has been used in the architectural discipline and building codes in defining minimum planning standards in different countries, especially when it comes to affordable housing [63]. A room is deemed feasible if it fits a certain number of predefined furniture blocks. However, the planning standards are only visual guides meant for manual placement of furniture blocks by architecture professionals and are not automated digital procedures. Inspired by the spatial scoring system developed by the City of Berlin's public housing provider [26] and the City of London's planning standard [64] we translate the manual workflow to an automated digital approach and procedurally place furniture blocks (Figure 7a) into a floor plan, where furniture blocks are placed recursively along the boundary geometry of each room (Figure 7b). By grouping furniture items inside a program together we can provide different simple configurations using a faster, less computationally intensive, procedural method.

Each apartment has a minimal number of furniture items that need to fit, to be a valid floor plan (Supplemental Figure 7). In the case of bedrooms and bathrooms we distinguish between a primary room, such as a bathroom with a bathtub and secondary bathroom, with toilet and sink only, in the larger apartments. We used the same minimal furniture to assess floor plans of Zurich, Singapore, and New York. The workflow is very flexible and could be adjusted to include more nuanced cultural requirements. An example floor plan subdivision is valid if all required furniture can be placed (Figure 7h). If the furniture placement is infeasible (Figure 7i) that is an indication that the hypergraph subdivision did not create a feasible layout.

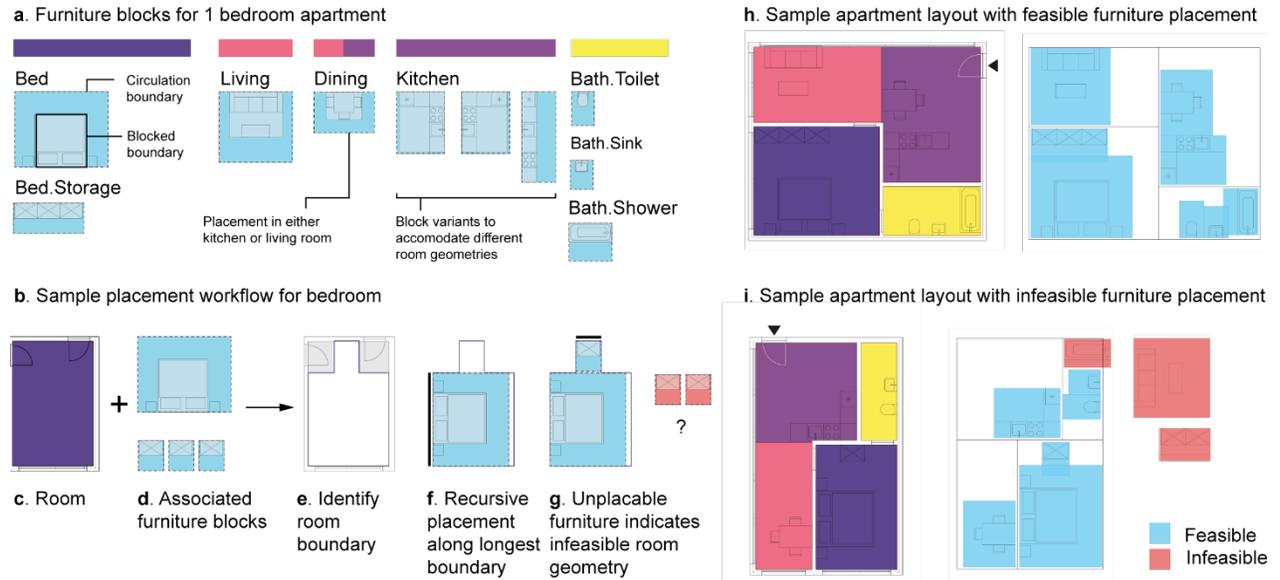

*Figure 7: Example floor plans for a one-bedroom apartment, where all required furniture can be placed (a), creating a valid apartment layout. If furniture is not placeable inside the room geometry an invalid apartment layout was created (b) and a different hypergraph should be chosen to subdivide the floor plan boundary.*

## 5.10. Excess area and emissions

To derive the excess carbon from excess area we use a floor plan furnished with a minimum furniture area. After placement of the furniture, we sum up the total furniture area and compare it to the minimum furniture area of the corresponding apartment size (Supplemental Figure 7). We derive the total excess area with a subdivision of the furniture area from the total apartment area and the carbon emissions from excess space by multiplying the excess area with the local grid carbon content and EUI (Equation 3-6). The emission difference of excess area and envelope upgrade $\Delta_e$ is derived from Equation 7, using the EUI results of the environmental simulation. If the emission delta $\Delta_e$ is positive the emissions from excess space exceed the emissions that could have been saved through a high-performance building envelope.

$$F_{tot} = \sum_{n=1}^{n}(F_n)$$

*Equation 3: $F_{tot}$ is the total furniture area (in m²) sum of all furniture areas $F_n$ of all furniture objects inside the apartment (extra rooms count as furniture, foyer rooms do not). If the $F_{tot}$ is smaller than the minimum furniture area (Supplemental Figure 7). If the furnishing was unsuccessful and $F_{tot}$ is clamped at the minimum furniture area.*

$$A_e = A_{apt} - (F_{tot} * M)$$

*Equation 4:* $A_e$ is the excess area (in m²) derived from subtracting the sum of all furniture areas $F_{tot}$ from the total apartment area $A_{apt}$ with a multiplier buffer. A positive $A_e$ indicates excess area (an apartment exceeding sufficiency), a value close to 0 indicates a good fit and a value of less than 0 indicates no excess area and a compact apartment. The multiplier (M) can be adjusted to cultural contexts. We use M=1.6 to create apartments with target areas according to the German public housing standard *[26]*: Studio 34m², 1 Bed 53.6 m², 2 Bed 72.6 m², 3 Bed 93.1 m², 4 Bed 105.9 m², 5 Bed 118.7 m².

$$C_e = A_e * EUI_s * g_{cc}$$

*Equation 5:* $C_e$ is the excess carbon emitted from an apartment per annum (kgCO2e/a), where $A_e$ is the excess area (m²) (*Equation 4*), $EUI$ the Energy Use Intensity (kWh/m²/a) derived from the energy simulation of the apartment with standard building envelope, and $g_{cc}$ the local grid carbon content (kgCO2e/kWh).

$$\Delta_e = C_e - (A * EUI_s * g_{cc} - A * EUI_{hp} * g_{cc})$$

*Equation 6:* $\Delta_e$ is the difference between the carbon emitted from an apartment from excess space $C_e$ (*Equation 5*), and the excess carbon emitted from not having an envelope upgrade, where $A$ is the apartment area (m²), $EUI_s$ the Energy Use Intensity (kWh/m²/a) of the apartment with standard envelope and $EUI_{hp}$ the Energy Use Intensity (kWh/m2/a) of the apartment with high performance building envelope, and $g_{cc}$ the local grid carbon content (kgCO2e/kWh)

# A hypergraph model shows the carbon reduction potential of effective space use in housing

Ramon Elias Weber*[a], Caitlin Mueller[a], Christoph Reinhart[a]

[a] Building Technology Program, Department of Architecture, Massachusetts Institute of Technology, 77 Massachusetts Avenue, Cambridge, MA 02139, USA, *reweber@mit.edu

## 1. Supplementary information

### 1.1. *Supplementary note on the hypergraph method*

The following supplemental figures, contain additional illustrations to explain the hypergraph method and its limitations. As shown in Supplemental Figure 1, the geometric subdivision can be applied to either retain area ratios or actual areas of the original floor plan, that the hypergraph is based on. For the purposes of this research and the automatic generation of new floor plans the ratio retention method was chosen. In order to make sure that rooms have approximate sizes similar to their original floor plan, only hypergraphs with an original area foot print within a 20% range were used.

Supplementary Figure 2 shows the application of hypergraphs on an apartment boundary that results in an infeasible floor plan layout. Three main failure cases were identified that for successful automated floor plan generation have to be discarded. We outline fast heuristics to discard such infeasible floorplans in "8.2.Supplementary note on heuristics" and "5.7.Apartment validity heuristic".

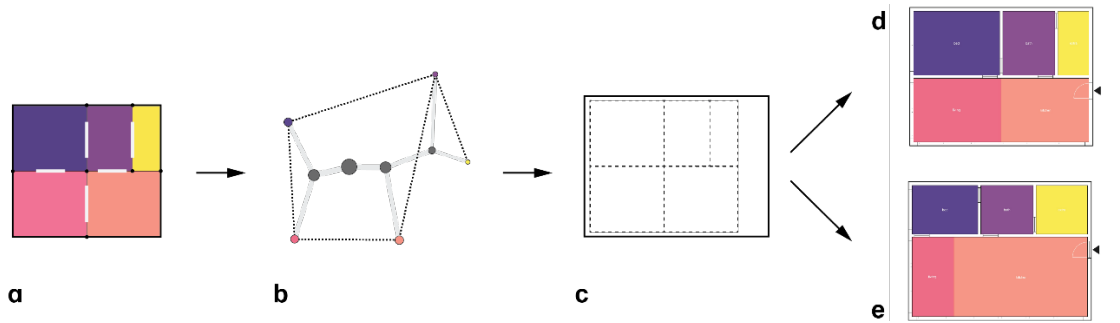

*Supplemental Figure 1: A floor plan (a) creates a hypergraph (b) and is applied to a different boundary geometry (c). The new boundary geometry can be subdivided to retain the ratios of the original floor plan (d) or the actual areas of the reference floor plan (e).*

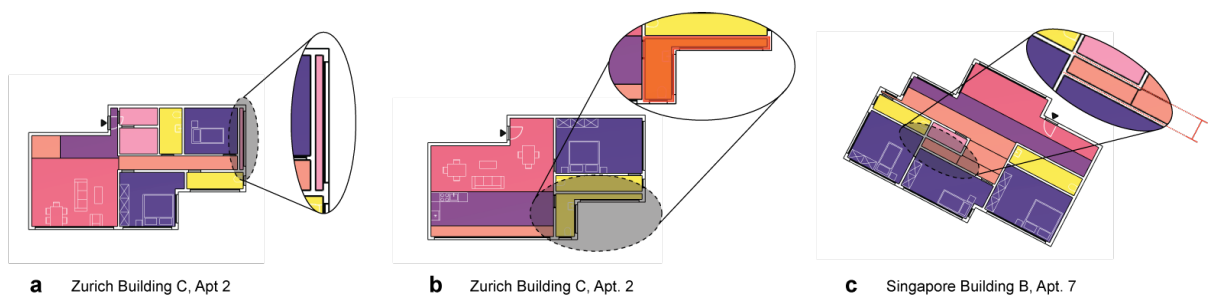

*Supplemental Figure 2: Different failure cases where the subdivision algorithm creates a geometrically valid but spatially infeasible floor plan: An infeasible room blocking a bedroom façade access (a), subdivision resulting in infeasible room geometry (b), and subdivision resulting in foyer spaces that are too thin to be passable (c).*

## 1.2. *Supplementary note on heuristics*

For the automatic generation of floor plans via the hypergraph method a series of geometric heuristics were developed that are fast to compute and help discard infeasible geometries. Supplemental Figure 3 illustrates the perimeter difference score δp, described in Equation 1. The perimeter difference score allows a quick assessment of the differences between two geometric shapes based on their boundary length. As shown in Supplemental Figure 4, this can be applied to all rooms in a given floor plan, creating a perimeter score of the source plan (that the hypergraph was derived from) and the resulting floor plan that the hypergraph was applied to. To cull floor plans with infeasible interior layout subdivisions the authors suggest a perimeter difference score of 0.1 and lower.

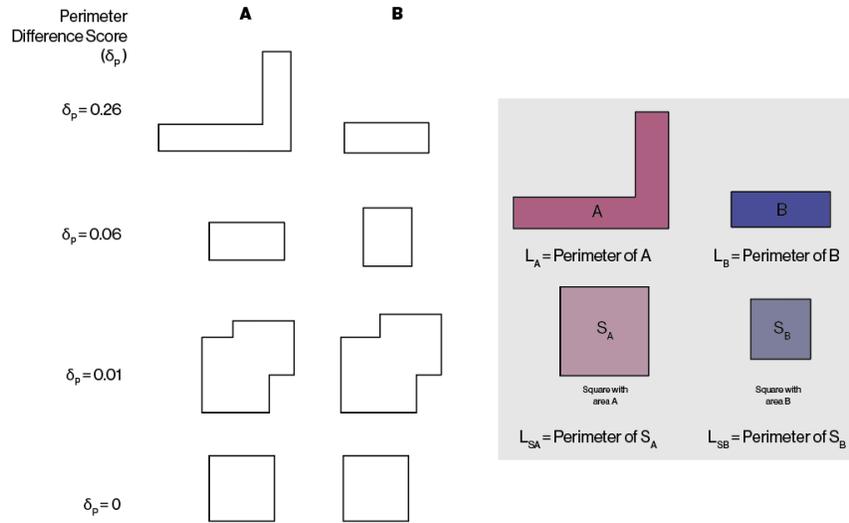

*Supplemental Figure 3: Example reference (A) and target (b) boundary room polygons with corresponding perimeter difference scores.*

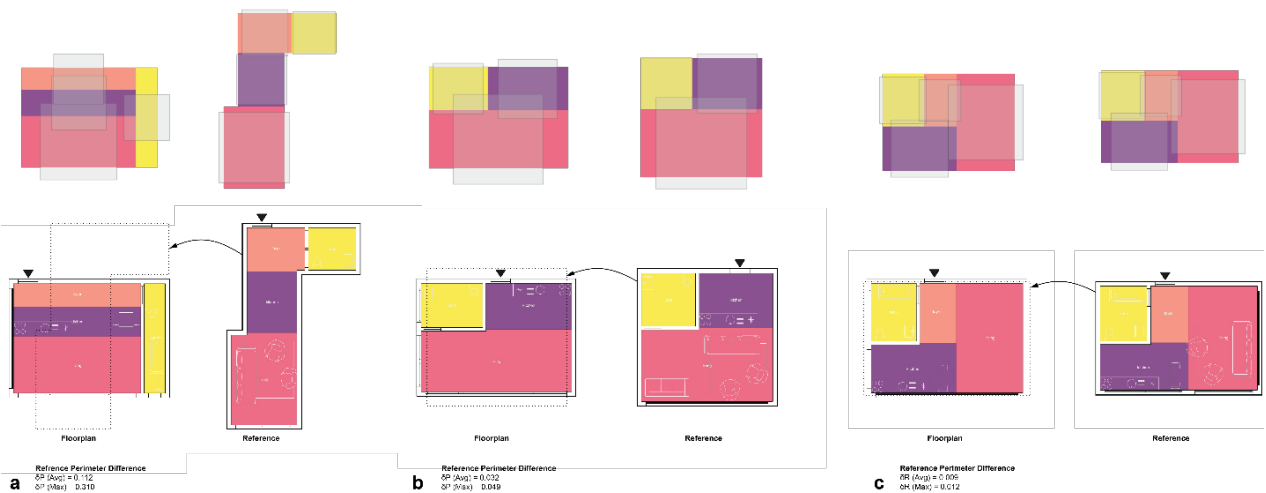

*Supplemental Figure 4: Example fitted floor plans, their reference floor plans and the corresponding perimeter difference $\delta_R$. A low fit with a value of $\delta_R$ (**average**) = 0.112 (a), a medium fit with a value of $\delta_R$ (**average**) = 0.08 and a very close fit with a value of $\delta_R$ (**average**) = 0.009.*

## 1.3. *Supplementary notes on environmental simulation*

In the scope of this research the daylight simulations are only conducted for rooms that are in need of daylight, based on program. As illustrated in Supplemental Figure 5 this includes living, kitchen, bed and foyer spaces and does not include bath or extra spaces.

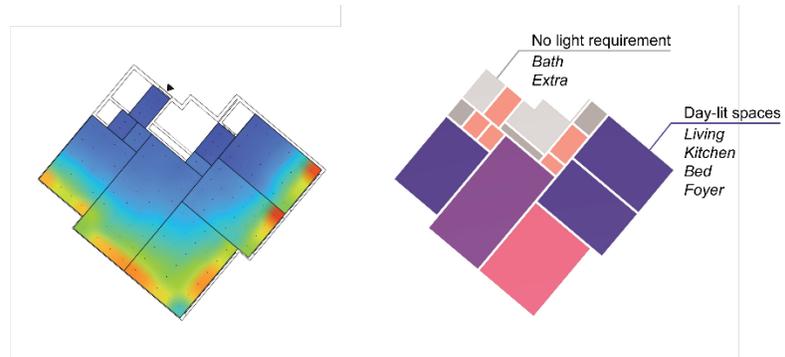

*Supplemental Figure 5: Example of an sDA calculation of apartment visualizing mean illuminance (left) and their corresponding program (right).*

## 1.4. *Supplementary notes on furniture placement*

Supplemental Figure 6 extends the description of section "5.9. Furniture placement" and shows a sample placement of bedroom furniture in full detail along the interior boundary of a room. Supplemental Figure 7 outlines the minimal furniture requirements of different apartment sizes.

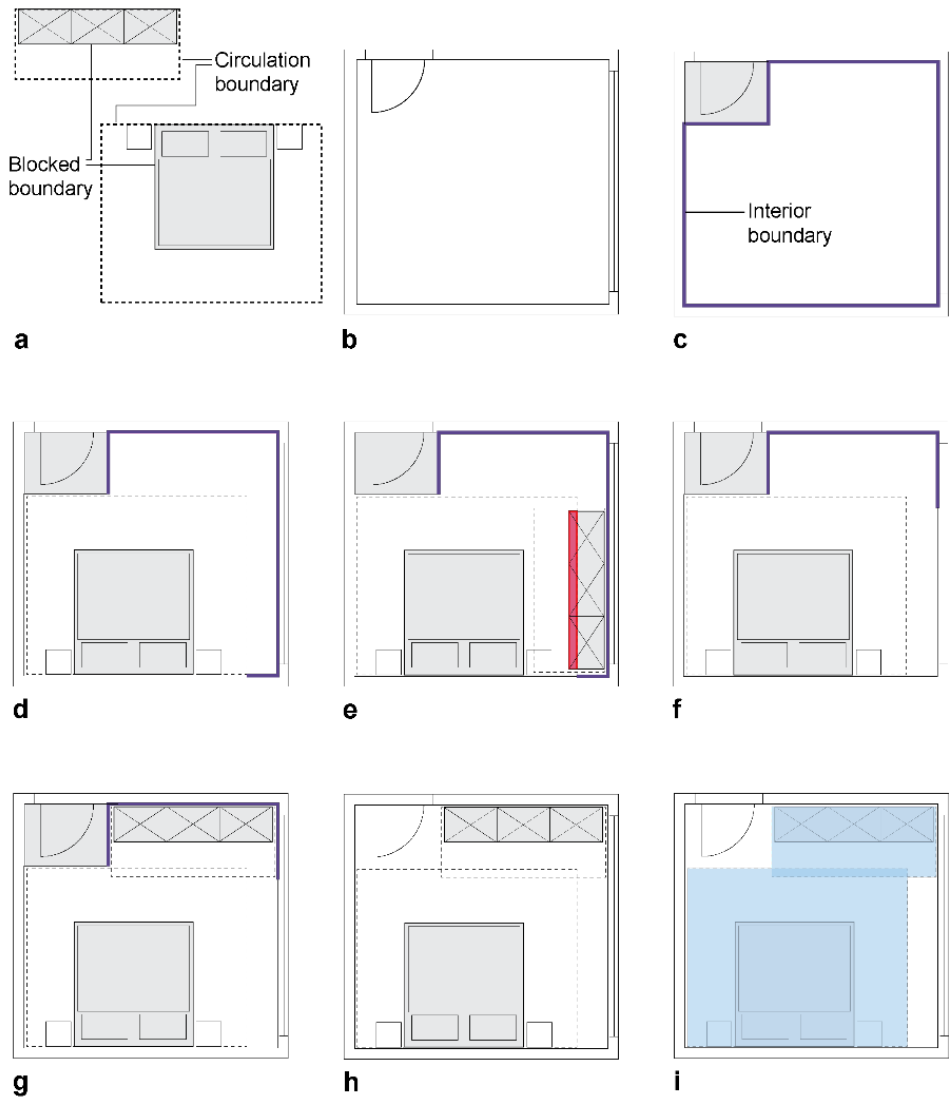

*Supplemental Figure 6: Procedure of furniture placement inside a sample room. Furniture objects are defined with a circulation boundary that is allowed to intersect with other circulation boundaries and blocked boundary that is not allowed to intersect with any object (a). A bed and drawer are placed in the bedroom (b). The circulation area of the door is subtracted from the interior boundary to create the interior placement polygon (c). The first furniture item (the bed) is placed along the first edge of the interior placement polygon (d) after successful placement the circulation bounds are subtracted from the interior boundary. The second furniture item (the drawer) is placed along the next line segment of the interior boundary. The blocked boundary of the drawer intersects with the circulation boundary of the bed (e), which is an infeasible placement. The interior boundary is shortened by the circulation boundary of the drawer (f), and in a second try the drawer is placed on the next line of the interior boundary polygon (g). A successful placement is achieved (h) and the furniture area of the room can be calculated (i).*

**a.** Studio apartment (minimum furniture area 21.4 m²)

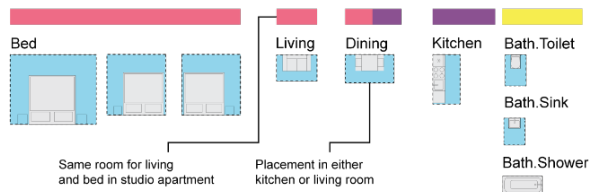

**b.** 1 Bedroom apartment (minimum furniture area 33.5 m²)

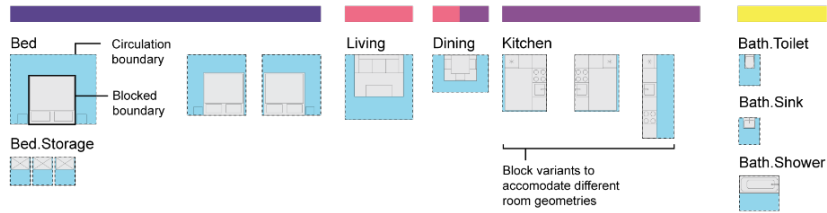

**c.** 2 Bedroom apartment (minimum furniture area 45.4 m²)

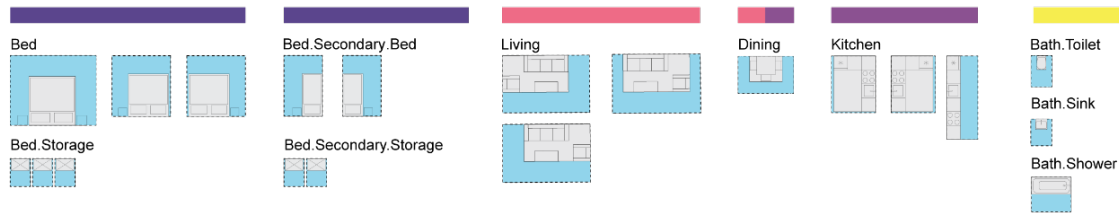

**d.** 3+ Bedroom apartment: (minimum furniture area 58.2 m² for 3 bed, 66.2 m² for 4 bed, 74.2 m² for 5 bed)

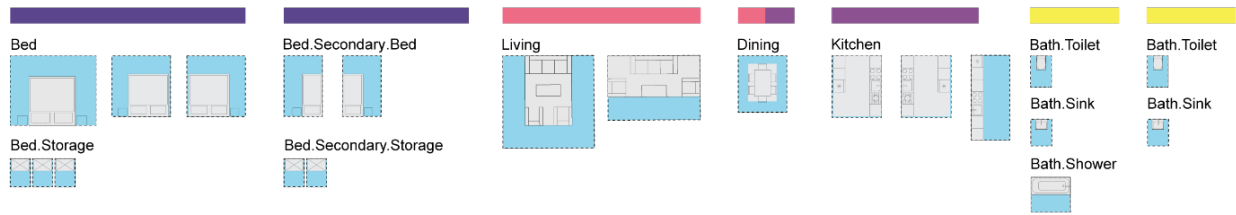

*Supplemental Figure 7: Minimum furniture blocks by apartment size*

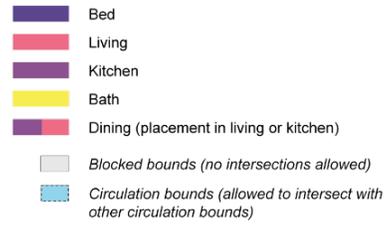

## 1.5. *Supplementary note on hypergraph samples*

Supplementary Figure 8 shows different examples of artificially generated layout subdivisions using the hypergraph method. The hypergraph subdivision is oriented to match the orientation of the circulation, resulting in different floor plan layouts.

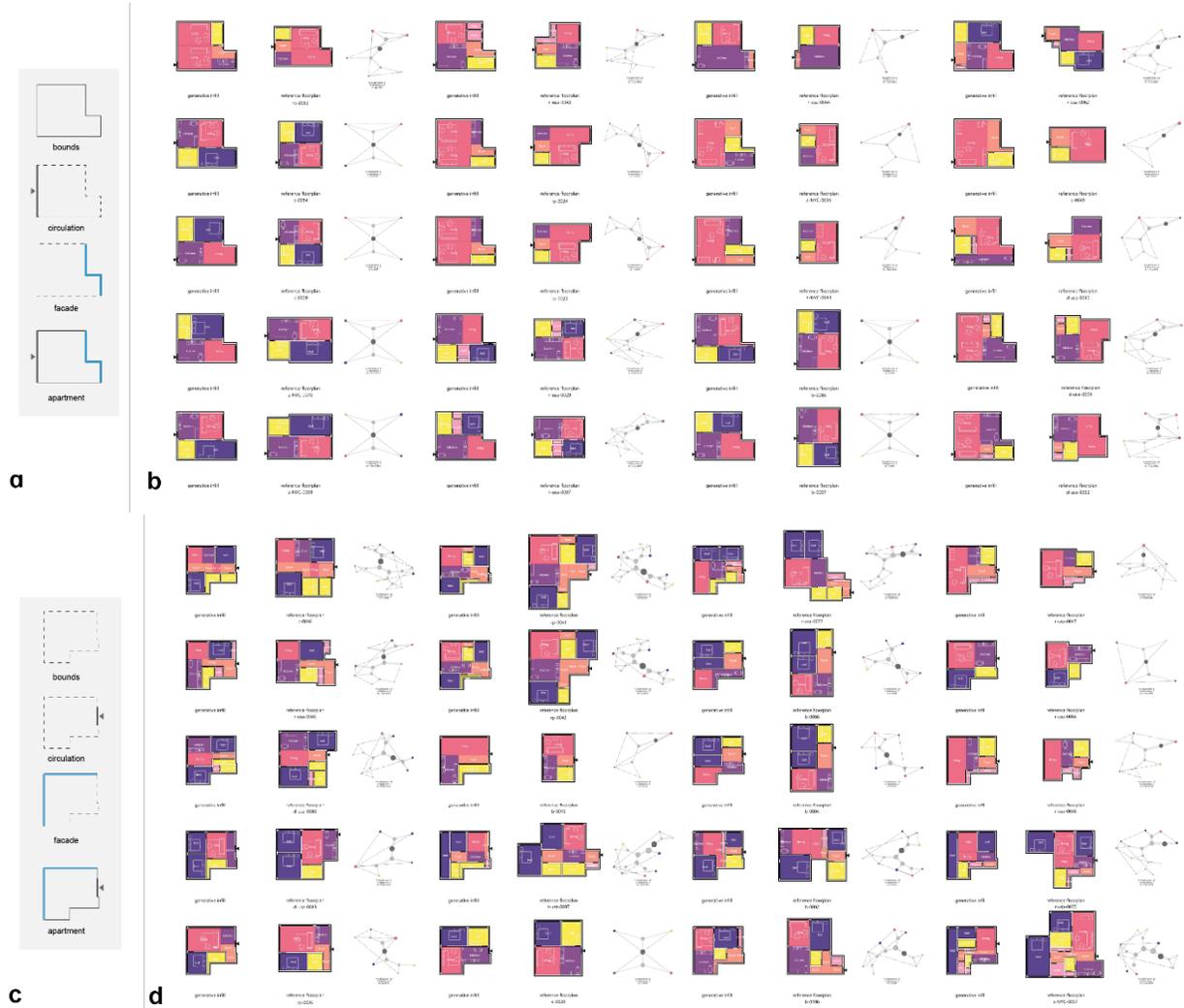

*Supplemental Figure 7: Input boundaries (a,c) with applied generative infill (b,d), where from left to right the resulting floor plan, the source floorplan and its corresponding hypergraph are shown.*

### 1.6. *Supplementary note on hypergraph PCA analysis*

To compare the hypergraphs, a principal component analysis (PCA) is conducted. As hypergraphs have different numbers of nodes and edges they can't be directly used for comparison. Instead, key attributes of the graphs of different lengths are converted into a one-dimensional vector. Supplemental Figure 9 outlines different attributes of the hypergraph that can be extracted and that capture the differences between different hypergraphs.

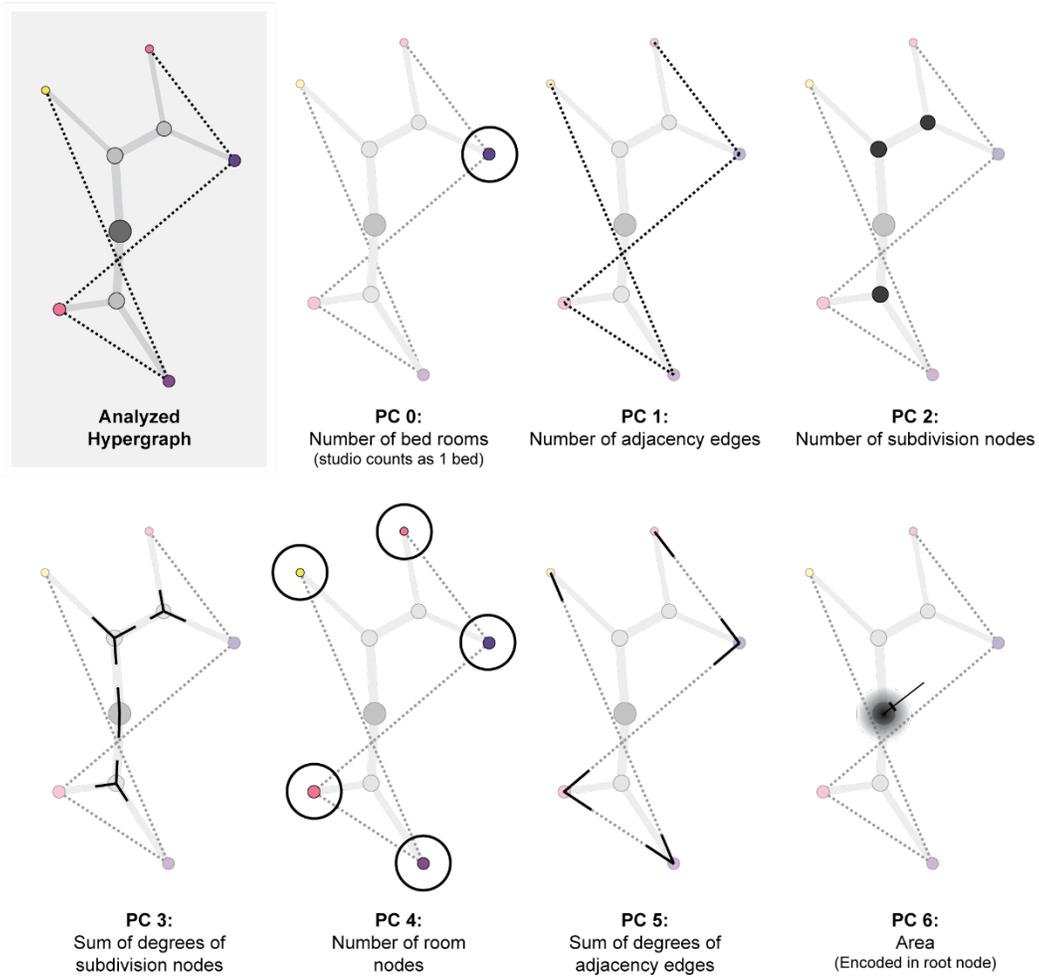

*Supplemental Figure 9: Visualization of graph properties that were used for inputs into PCA.*

## 1.7. *Supplementary note on artificially generated floor plans*

For testing of the hypergraph method to generate artificially generated floor plans for real world buildings a dataset of 12 buildings from Zurich, Singapore and New York was created. The buildings have been anonymized and are shown in

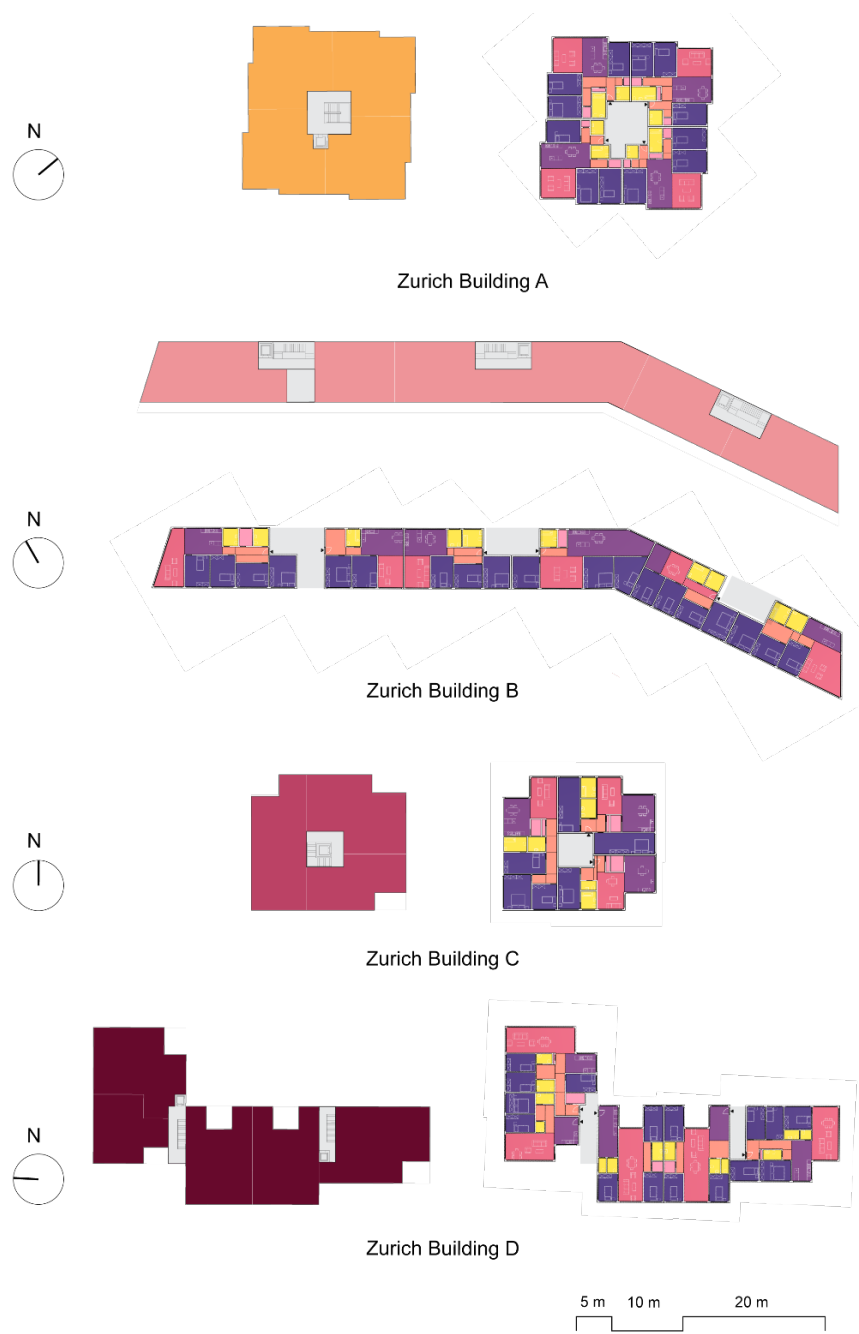

*Supplemental Figure 10: Zurich Building A-D*

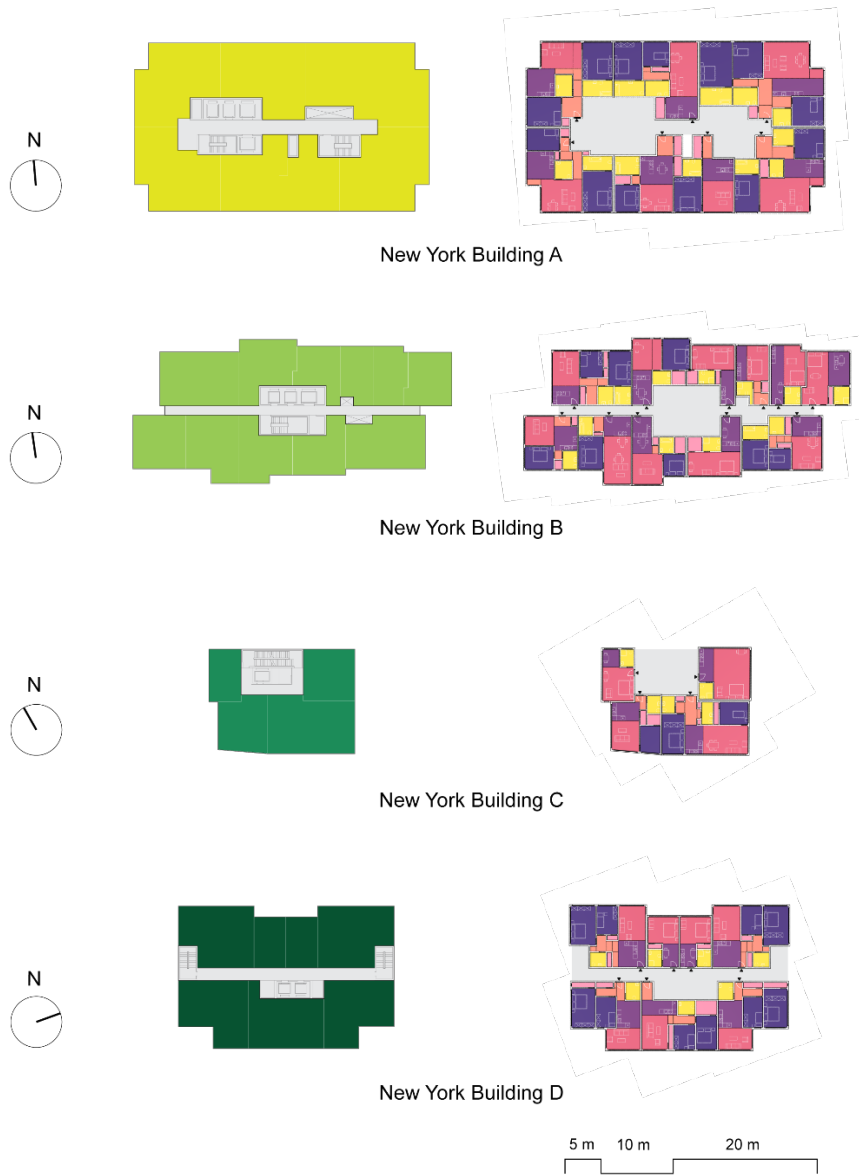

*Supplemental Figure 11: New York Building A-D*

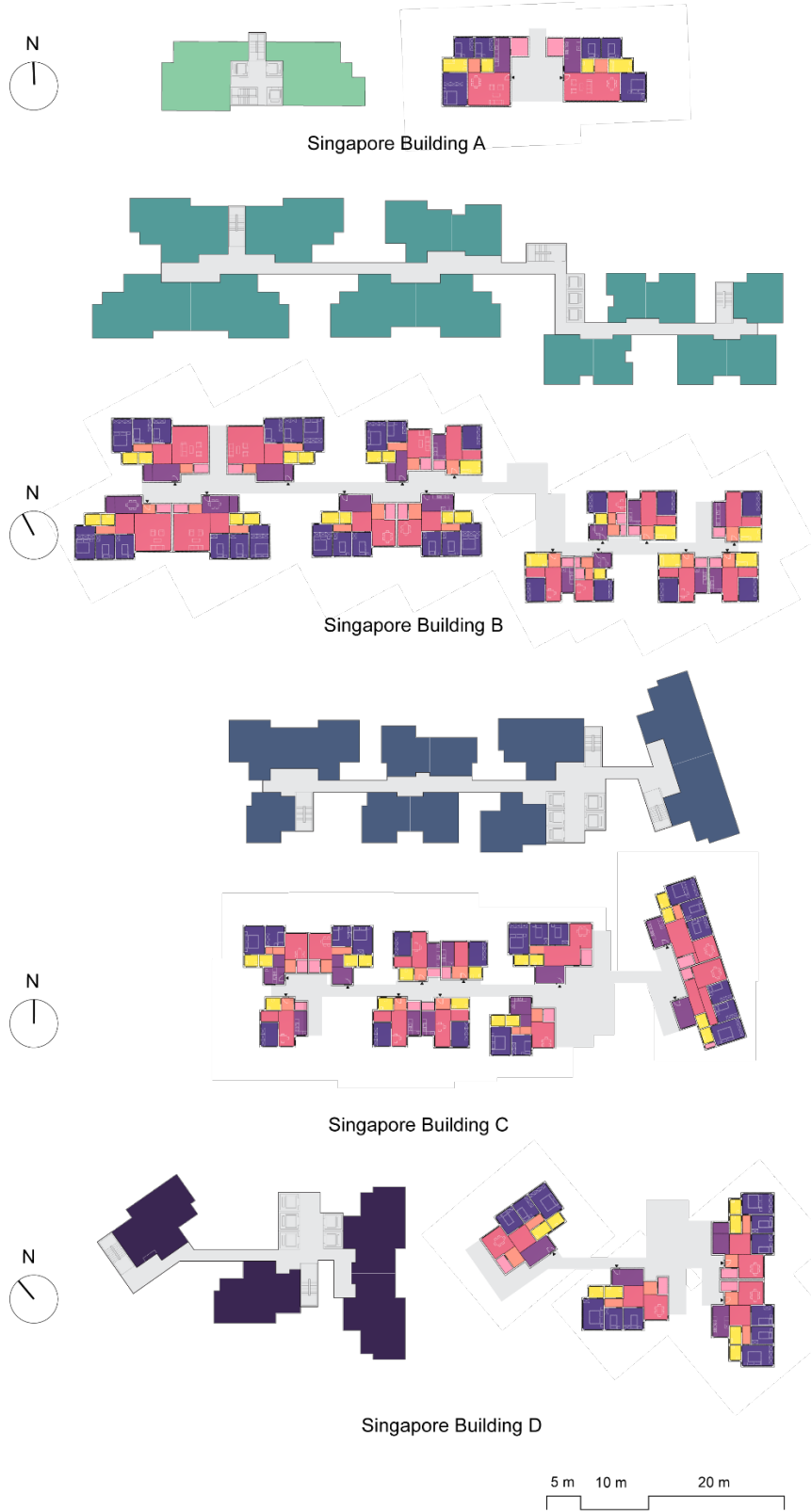

*Supplemental Figure 12: Singapore Building A-D*